\documentclass[11pt,aps,onecolumn,preprintnumbers,floatfix,superscriptaddress]{revtex4-2}

\usepackage[final]{graphicx}
\usepackage{hyperref}
\usepackage{amsmath}
\usepackage{bbm}
\usepackage{amsfonts}
\usepackage{amssymb}
\usepackage{latexsym}
\usepackage{graphicx}
\usepackage[english]{babel}
\usepackage{multirow}
\usepackage{float}
\usepackage{url}
\usepackage{slashed}
\usepackage{xcolor}
\usepackage[utf8]{inputenc}
\usepackage{verbatim}
\usepackage{lipsum}

\newcommand{\beq}{\begin{equation}}
\newcommand{\eeq}{\end{equation}}
\newcommand{\bi}{\begin{itemize}}
\newcommand{\ei}{\end{itemize}}

\newcommand{\be}{\begin{equation}}
\newcommand{\ee}{\end{equation}}
\newcommand{\ba}{\begin{array}}
\newcommand{\ea}{\end{array}}
\newcommand{\bea}{\begin{eqnarray}}
\newcommand{\eea}{\end{eqnarray}}

\makeatletter
\renewcommand{\maketag@@@}[1]{\hbox{\m@th\normalsize\normalfont#1}}%
\makeatother

\newlength{\halfpagewidth}
\divide\halfpagewidth by 2

\begin{document}

\title{Baryogenesis via the Chiral Magnetic Effect in a First-Order Electroweak Phase Transition}

\author{Hui Liu}
\email{lh001@stu.cqu.edu.cn}
\affiliation{Department of Physics and Chongqing Key Laboratory for Strongly Coupled Physics, Chongqing University, Chongqing 401331, P. R. China}

\author{Renhui Qin}
\email{20222701021@stu.cqu.edu.cn}
\affiliation{Department of Physics and Chongqing Key Laboratory for Strongly Coupled Physics, Chongqing University, Chongqing 401331, P. R. China}

\author{Ligong Bian\footnote{Corresponding Author.}}
\email{lgbycl@cqu.edu.cn}
\affiliation{Department of Physics and Chongqing Key Laboratory for Strongly Coupled Physics, Chongqing University, Chongqing 401331, P. R. China}

\begin{abstract}
  
    In this paper, we investigate the generation of the baryon asymmetry of the universe during the first-order electroweak phase transition.  We first study the generation of the helical magnetic field in the framework of the standard model effective field theory with a CP-violating operator. We show that, when the chiral magnetic effect is absent, the helical magnetic field and effective chemical potential cannot generate enough baryon asymmetry when vacuum bubbles collide. We further find that the chiral magnetic effect can amplify the lepton asymmetry in the early universe during the phase transition. We present the baryon asymmetry interpretation requirement on certain parameter spaces of the phase transition and the primordial magnetic field.

\end{abstract}
\maketitle
\tableofcontents
\date{\today}
\section{Introduction}

The origin of the baryon asymmetry in the present Universe is one of the biggest problems in cosmology and particle physics. The prevailing belief is that generating baryon asymmetry within the Standard Model (SM) of particle physics is almost impossible because it is hard to provide the departure from a thermal equilibrium environment, i.e., one of the three.
Sakharov’s conditions~\cite{Sakharov:1967dj}. 
First-order electroweak phase transition (EWPT) can lead to the departure of the thermal equilibrium, which is generally predicted in many new physics models indicating the discovery of new physics beyond the SM, such as SM extended with a dimensional-six operator $(\phi^{\dagger}{\phi})^3/\Lambda^2$ with $\Lambda$ being the new physics (NP) scale~\cite{Grojean:2004xa, Grojean:2006bp}, xSM~\cite{Zhou:2019uzq, Bian:2019bsn, Alves:2018jsw, Jiang:2015cwa}, 2HDM~\cite{Cline:2011mm, Dorsch:2013wja, Dorsch:2014qja, Bernon:2017jgv, Andersen:2017ika, Kainulainen:2019kyp}, George-Macheck model~\cite{Zhou:2018zli}, and NMSSM~\cite{Bian:2017wfv, Huber:2015znp}. Besides the generation of stochastic gravitational waves (GWs)~\cite{Mazumdar:2018dfl, Caprini:2015zlo, Caprini:2019egz} in a first-order phase transition, it has also been proposed to generate magnetic fields (MFs) that may seed cosmological MFs~\cite{Durrer:2013pga, Subramanian:2015lua, Vachaspati:2020blt, Di:2020kbw}\footnote{The 
Temporary charge-breaking phase in the early universe may affect the MFs generation~\cite{
Basler:2024aaf, Aoki:2023lbz}. }, which have been observed through extensive astronomical observations~\cite{Xu:2021kwb, Han:2002mz}. Refs~\cite{Kibble:1995aa, Ahonen:1997wh} studied the generation of ring-like MFs arising from the collisions of bubbles in an Abelian Higgs model. 
Refs.~\cite {Stevens:2007ep, Stevens:2009ty} further studied the MFs production during the EWPT process. 

Many studies have investigated the connection between a primordial magnetic field (PMF)
And the baryon asymmetry of the Universe (BAU). Previously, the generation of baryon asymmetry in the universe has been extensively studied by hypothesizing the existence of an observable magnetic field as an initial condition in the early universe. Early studies show that a helical hypermagnetic field can arise in the symmetric phase of the EW plasma due to a preexisting lepton asymmetry carried by right-chiral electrons~\cite{Joyce:1997uy, Campbell:1992jd}, and the 
Preexisting stochastic hypermagnetic field would induce the generation of baryon number isocurvature fluctuations~\cite{Giovannini:1997eg, Giovannini:1997gp, Giovannini:1999by, Giovannini:1999wv}. 
Later, more detailed studies revealed the relationship between helical magnetic fields in the early Universe and lepton asymmetry after taking into account the evolution of magnetic fields~\cite{Boyarsky:2011uy, Boyarsky:2012ex, RostamZadeh:2015xnd, Boyarsky:2015faa, Semikoz:2003qt, Semikoz:2004rr, Semikoz:2005ks, Semikoz:2009ye, Akhmetev:2010qdw, Dvornikov:2011ey, Semikoz:2012ka, Dvornikov:2012rk, Semikoz:2013xkc, Semikoz:2015wsa, Semikoz:2016lqv}. 
 Ref.~\cite{Pavlovic:2016mxq} modified magnetohydrodynamics (MHD) after EWPT to consider the evolution of lepton asymmetry.

 In a magnetized chiral plasma, an additional non-dissipative matter current arises along the direction of the magnetic field. This phenomenon is referred to as the chiral magnetic effect (CME) ~\cite{Vilenkin:1980fu, Kharzeev:2004ey, Kharzeev:2007tn, Kharzeev:2007jp}. In the symmetric phase, the matter chiral magnetic current for a chiral fermion arises due to the CME. The CME, together with the $U(1)_Y$ Abelian anomaly, establishes a connection between the hypermagnetic helicity and the fermion number densities~\cite{Joyce:1997uy, Giovannini:1997eg, tHooft:1976rip}.  Previous studies show that the BAU 
 can be generated before the EWPT epoch~\cite{Fujita:2016igl, Abbaslu:2021mkt, Semikoz:2013xkc, Domcke:2022uue, Semikoz:2016lqv, Kamada:2016eeb, RostamZadeh:2015xnd, Anber:2015yca}\footnote{For the situation with the effects of gravity, we refer to Refs.\cite{Kushwaha:2020nfa, Kushwaha:2021csq}}, and Ref.~\cite{Kamada:2016cnb} performs a more systematic study after taking into account the electroweak crossover effect.
 
When a dimension-six CP-violating operator appears, we first study the relation between the generated MFs and the baryon asymmetry. The baryogenesis here solely comes to the relation between the generated helical magnetic fields and the change of Chern-Simons number~\cite{Vachaspati:2001nb}. We give the relation between the helicity of the generated MF and the new physics scale of the CP violation, and show that the CME effect is essential for the baryon number generation. Then, we present the framework to generate baryon asymmetry during the first-order EWPT, and then study how to obtain the observed BAU in the appearance of the CME effect. We consider the system of extended MHD equations after considering the chiral anomaly~\cite{Giovannini:1997eg}. 

The paper is organized as follows. In Section.III, the generation of the MF and its property through the collision of two bubbles during the first-order EWPT is presented as a concrete illustration. In Section.II, we quantitatively estimate the PT parameters and the required magnetic field strength required by the explanation of the observed BAU. In Section.IV, the cosmological observation of the magnetic field is given. And then we conclude with a summary and discussion in the Section. V. In the Appendix, we provide some complementary details.

\section{Baryogenesis without CME}

In this section, employing the Standard Model extended by two dimension-six operators, i.e., $O_6=(\Phi^\dag \Phi)^3/\Lambda^2$ and $\tilde{O}_{\Phi B}=\frac{\Phi^{\dagger}\Phi}{\Lambda
_{CPV}^2}B_{\mu\nu}\tilde{B}^{\mu\nu}$, we first study the scenario of magnetic field production and the following baryon asymmetry generation. 
Where the first CP even operator can provide a first-order EWPT when $\Lambda\sim[600,900]$ GeV~\cite{Qin:2024idc} and the second CP-violating (CPV) operator seeds the helical magnetic field to induce the Chern-Simon numbers variation and the baryon asymmetry generation, in the following study, we take $\Lambda=\Lambda_{CPV}$. 
The two dimension-six operators modify the equations of motion for the Higgs and isospin gauge fields, see Appendix~\ref {sec:eom} for details. The system under study has an $O(2)$ symmetry in the spatial coordinate. We follow the analysis in Ref.~\cite{Yang:2021uid} and express the EOM in a coordinate $(\tau,z)$ which has an O(2) symmetry. 

\begin{figure}[!htp]
\centering
\includegraphics[width=0.4\linewidth]{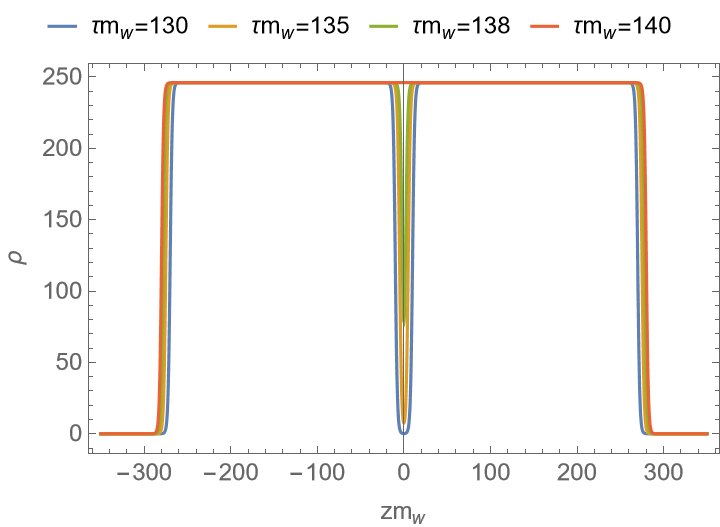}
\includegraphics[width=0.4\linewidth]{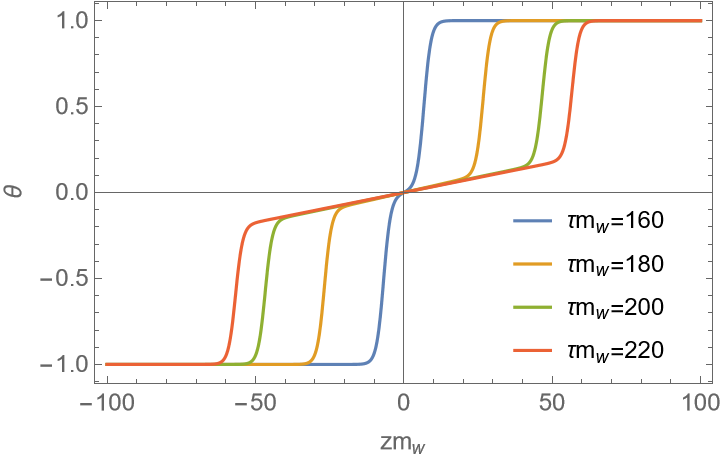}
\caption{Left: The scalar field of two colliding bubbles in the case of $t_{col}m_{\rm w}=200$; Right: The Higgs phase $\Theta$ is the field is shown as a
The function of distance z along the axis of collision for $\tau m_{\rm w}=160,180,200,220$, with $\Theta_0=1$.}
\label{Twobubbles}
\end{figure}

The Higgs doublet in the Unitary gauge takes the form of
\beq
\label{phi}
         \Phi(x)  = \left( \begin{array}{clcr} 0 \\
                                \phi(x)
         \end{array} \right) \; ,
\eeq
To consider the effects of bubble collision during the MF production, we use the definitions of
\begin{equation}
\begin{aligned}
&\phi(x) \equiv \rho(x)e^{i\Theta(x)},  \\
&|\phi(x)|^2 = \rho(x)^2 .
\label{phirho}
\end{aligned}
\end{equation}
where $\Theta(x)$ is the phase of the Higgs field, and $\rho(x)$ is its magnitude. We here illustrate the MF generation from bubble collisions by employing the two-bubble system. To obtain the magnetic field generated by two bubble collisions, we solve the equation of the phase $\Theta$ in the coordinate $(\tau,z)$:
\beq
\frac{2}{\tau}\frac{\partial \Theta}{\partial \tau}+\frac{\partial^2 \Theta}{\partial^2 \tau}-\frac{\partial^2 \Theta}{\partial z^2}+\frac{2}{\rho(x)}\frac{\partial \rho(x)}{\partial \tau}\frac{\partial \Theta}{\partial \tau}
-\frac{2}{\rho(x)}\frac{\partial \rho(x)}{\partial z}\frac{\partial \Theta}{\partial z}=0\;.
\label{eqtheta}
\eeq
 We assume that before the bubble collision occurs, the Higgs phase for a single bubble is constant throughout the bubble, and the two bubbles have different phases. the boundary conditions of $\Theta$ are given by
\begin{equation}
\begin{aligned}
&\Theta(\tau=t_{col},z)=\Theta_0\epsilon(z),~~
\frac{\partial}{\partial\tau}\Theta(\tau=t_{col},z)=0.
\end{aligned}
\end{equation}
where $\epsilon(z)$ is the sign of $z$ and $0<\Theta_0<\pi/2$ is the initial Higgs phase in one of the colliding bubbles.
Here, we consider the simplest case: two identical bubbles nucleate simultaneously.
We define $\rho(x)$ as follows for such a scenario,
\begin{equation}
\begin{aligned}
    \rho&=\frac{v}{2}(1-\tanh(\frac{|z-v_{\rm w} t_{col}|-\tau}{l_{\rm w}}))
    +\frac{v}{2}(1-\tanh(\frac{|z+v_{\rm w} t_{col}|-\tau}{l_{\rm w}})).
\end{aligned}
\label{rho2}
\end{equation}
where $\tau=\sqrt{v_{\rm w}^2t^2-r^2}$ with $r^2=x^2+y^2$, $r$ is the radius of the bubble in the $(x,y)$ plane. One bubble locates at $(t,x,y,z)=(0,0,0,v_{\rm w} t_{col})$, and the other one locates at the position of $(t,x,y,z)=(0,0,0,-v_{\rm w} t_{col})$, where $v_{\rm w}$ is bubble velocity, $t_{col}$ is collision time, and $l_{\rm w}$ is the thickness of a bubble. We suppose they are expanding with the same velocity $v_{\rm w}$ so that they will be in a collision at the position $z=0$, thus the collision time is $t_{col}$ and $v=246$ GeV is the expectation value of the Higgs vacuum. See Fig.~\ref{Twobubbles} for illustration of the solution of Eq.~(\ref{eqtheta}) accompanied with the $\rho$ described by Eq.~(\ref{rho2}) when $\Lambda=600$ GeV, here the bubble velocity has been calculated to be $v_{\rm w}=0.7$, and the bubble wall width at the nucleation is $l_{\rm w}=0.025/{\rm GeV}$, see Appendix~\ref{sec:csd}.

We solved Eq. (\ref{eqtheta}) numerically, and then from Eq. (\ref{jem}), the $j_\nu^{em}(x)$ take the form
\begin{equation}
    \begin{aligned}
        &j_\nu^{em}(x)=(j_z(\tau,z),x_{\alpha}j(\tau,z)).
    \end{aligned}
\end{equation}
Where 
\begin{equation}
    \begin{aligned}
        &j_z(\tau,z)=-(\frac{c_1}{g}+c_2)\frac{8\rho(x)((\nabla_z \times \vec E))}{\Lambda^2}      -\frac{g'}{c_2}\rho{(x)}^2\frac{\partial \Theta(\tau,z)}{\partial z},\\
        &j(\tau,z)=(\frac{c_1}{g}+c_2)\frac{8\rho(x)\frac{\partial \rho(x)}{\partial \tau}B}{\Lambda^2}       -\frac{g'}{c_2}\rho{(x)}^2\frac{1}{\tau}\frac{\partial \Theta(\tau,z)}{\partial \tau}\;,\nonumber
    \end{aligned}
\end{equation}
and $x_{\alpha}=(v_{\rm w}t,-x,-y)$. The electromagnetic field is similar in form to the electromagnetic current.
\begin{equation}
    \begin{aligned}
        &A_{\nu}^{em}=(a_z(\tau,z),x_{\alpha}a(\tau,z))\;.
    \end{aligned}
\end{equation}
In the axial gauge $a_z=0$, Maxwell's equations become~\cite{Stevens:2012zz}:
\begin{equation}
    \begin{aligned}
        &-\frac{\partial^2 a(\tau,z)}{\partial z^2}=j(\tau,z)\;.
    \end{aligned}
\end{equation}
By applying the boundary conditions, specifically: $a(\tau_0,z)=0$, and $\partial_za(\tau=0,z)=0$, we have
\begin{equation}
    \begin{aligned}
        &a(\tau,z)=-\int_{-\infty}^{z}dz^{\prime}\int_{-\infty}^{z^{\prime}}j(\tau,z'')dz^{\prime\prime}.
    \end{aligned}
\end{equation}
After applying $\bf B=\nabla \times {\bf A^{em}}$, we can get the MF,
 \begin{equation}
     \begin{aligned}
         &B_x=-y\int_{-\infty}^{z}j(\tau,z')dz',\\
         &B_y=x\int_{-\infty}^{z}j(\tau,z')dz',\\
         &B_z=0\;.\nonumber
     \end{aligned}
 \end{equation}
  Finally, we have determined the numerical value of the magnetic field $\sqrt{B_x^2+B_y^2+B_z^2}$ resulting from bubble collisions. For more details, we refer to Appendix~\ref{sec:eom}.

 \begin{figure}[!htp]
\centering
\includegraphics[width=0.4\linewidth]{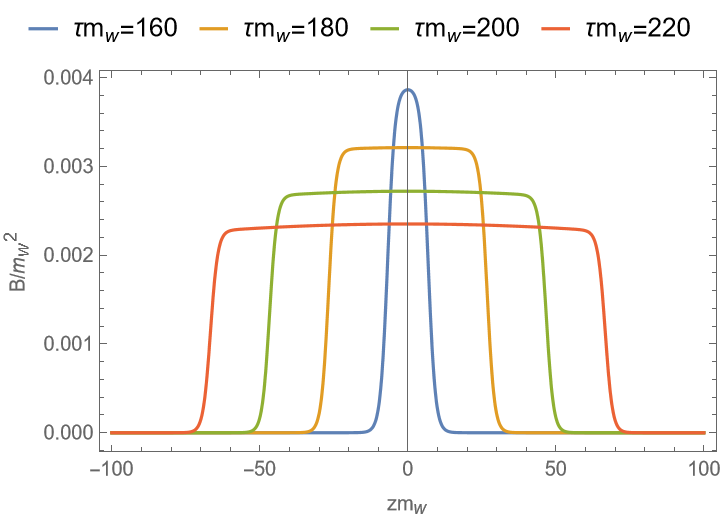}
\includegraphics[width=0.4\linewidth]{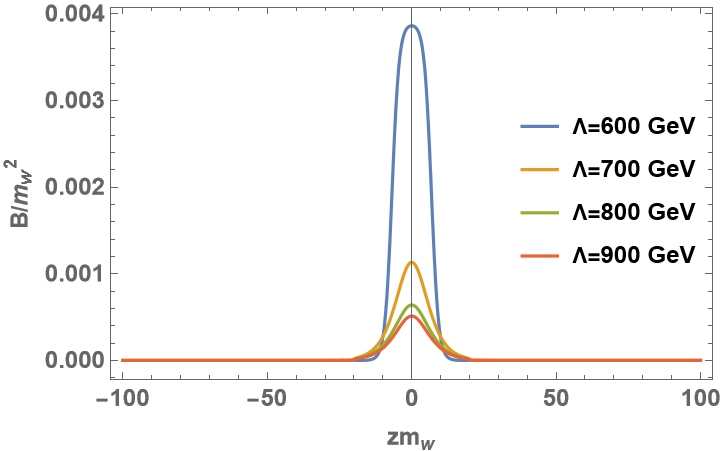}
\caption{ Left: The strength of the MFs as a function of the distance $z$ along the axis of collision at different times when $v_{\rm w}=0.7,~l_{\rm w}=0.025,~\Lambda=600$ GeV. Right: The strength of the MFs at $\tau m_{\rm w}=160$ for $\Lambda =600$ GeV,  $v_{\rm w}=0.33,~l_{\rm w}=0.048$ for $\Lambda =700$ GeV,  $v_{\rm w}=0.14,~l_{\rm w}=0.11$ for $\Lambda =800$ GeV,  $v_{\rm w}=0.07,~l_{\rm w}=0.19$ for $\Lambda =900$ GeV.}
\label{Bfields}
\end{figure}

\begin{figure}[!htp]
\centering
\includegraphics[width=0.6\linewidth]{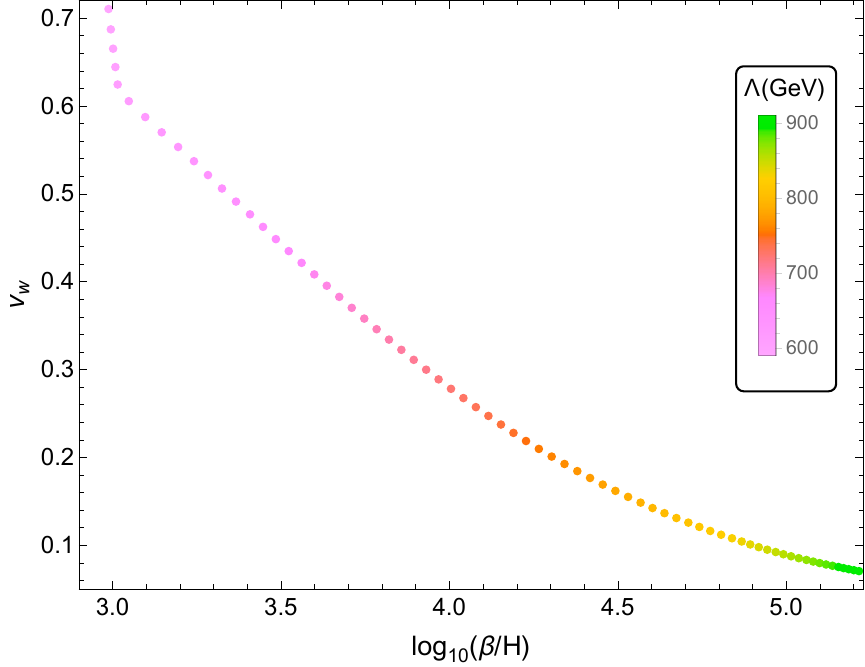}

\caption{ The bubble wall velocity and the phase transition duration for different NP scales.}\label{figvwbetah}
\end{figure}

The left panel of Fig.~\ref{Bfields} shows that the largest MF strength appears in the $z=0$ plane with $B\sim 0.001 m_W^2$. And, the right panel indicates that a lower cut-off scale $\Lambda$ with faster bubble wall velocity yields a higher magnitude of the MF strength. The parameters of 
$v_{\rm{w}}$ and $\beta/H$ are depicted in the left plot of Fig.~\ref{figvwbetah} for different NP scale $\Lambda$, where one can find that smaller NP scale can yield faster bubble wall velocity and slower PT, and therefore a larger chemical potential $\Delta\mu$ and a stronger MF are required to generate the observed BAU considering the CME effect aforementioned in the previous section. 

%From the dimension-six operator 
%\begin{equation}
%  \mathcal{L}_{\text{CPV}}
 % \;=\;
 % \frac{1}{\Lambda^2}\,\rho^2(x)\,
 % \frac{ g'^2}{32\pi^2}\,
 % B_{\mu\nu}\tilde{B}^{\mu\nu},
 % \label{eq:CPV-operator}
%\end{equation}
%where $\rho^2(x)$ from Eq.~\eqref{phirho},

%^Using Eq.~\eqref{eq:anomaly}, the CPV operator in Eq.~\eqref{eq:CPV-operator} can be rewritten as
%\begin{equation}
 % \mathcal{L}_{\text{CPV}}
 % \;=\;
  %\frac{1}{\Lambda^2}\,\rho^2\,\partial_\mu j_Y^\mu
  %\;=\;
  %-\,\frac{1}{\Lambda^2}\,
  %(\partial_\mu \rho^2)\, j_Y^\mu,
  %\label{eq:Lcpv-jY}
%\end{equation}
%where in the second step we integrated by parts and dropped a total derivative.
In the following, we demonstrate how to generate the baryon asymmetry from the generated MF after considering the relation between the Chern-Simons number and the MF helicity. Firstly, it is convenient to introduce the dimensionless background field $ \theta_{\rm eff}(x)\equiv\rho^2(x)/\Lambda^2$, 
then the CPV operator $\tilde{O}_{\phi B}$ can be written as
\begin{equation}
  \mathcal{L}_{\text{CPV}}
  \;=\;
  \theta_{\rm eff}(x)\,
  \frac{ g'^2}{32\pi^2}\,
  B_{\mu\nu}\tilde{B}^{\mu\nu}.
\end{equation}
Using the chiral anomaly equation
\begin{equation}
  \partial_\mu j_Y^\mu
  \;=\;
  \frac{ g'^2}{32\pi^2}\,
  B_{\mu\nu}\tilde{B}^{\mu\nu}\;,
  \label{eq:anomaly}
\end{equation}
one has
$
  \mathcal{L}_{\text{CPV}}
  \;=\;
  \theta_{\rm eff}\,\partial_\mu j_Y^\mu
  \;=\;
  -\,(\partial_\mu\theta_{\rm eff})\,j_Y^\mu$,
after the integration by parts. In a homogeneous background, with $\rho=\rho(t)$ and $j_Y^\mu \simeq (n_Y,\mathbf{0})$ in the plasma rest frame, the CPV contribution to the effective Lagrangian takes the form of
$\mathcal{L}_{\text{CPV}}
  \;\simeq\;
  -\,\mu_{\text{eff}}^Y(t)\,n_Y(t)$
Here, we adopt the effective Chern-Simons chemical potential~\cite{Garcia-Bellido:1999xos}:
$ \mu_{\text{eff}}^Y(t)
  \;\equiv\;
  \dot{\theta}_{\rm eff}(t)
  $.

%With this convention, the minus sign is kept in the Lagrangian, while the effective chemical potential $\mu_{\text{eff}}^Y$ enters with a positive sign in the modified Ohm's law and in the evolution of the hypermagnetic field.
In the MHD limit, the hypercharge current is given by the generalized Ohm's law~\cite{Joyce:1997uy, Semikoz:2013xkc, Kamada:2016eeb}
\begin{equation}
  \mathbf{J}_Y
  \;=\;
  \sigma_Y\bigl(\mathbf{E}_Y + \mathbf{v}\times\mathbf{B}_Y\bigr)
  + \frac{ g'^2}{8\pi^2}\,
    \mu_{\text{eff}}^Y(t)\,\mathbf{B}_Y,
  \label{eq:Ohm-mu}
\end{equation}
 According to Eq.~(\ref{amaxwell}), we solve for the hyperelectric field,
\begin{equation}
  \mathbf{E}_Y
  \;=\;
  \frac{1}{\sigma_Y}\left(
    \nabla\times\mathbf{B}_Y
    - \frac{ g'^2}{8\pi^2}\,
      \mu_{\text{eff}}^Y\,\mathbf{B}_Y
  \right)
  - \mathbf{v}\times\mathbf{B}_Y.
  \label{eq:Ey-from-Ohm-mu}
\end{equation}

Using Eq.~\eqref{eq:Ey-from-Ohm-mu} and neglecting the $\mathbf{v}\times\mathbf{B}_Y$ contribution to $\mathbf{E}_Y\cdot\mathbf{B}_Y$, we find
\begin{equation}
  \mathbf{E}_Y\cdot\mathbf{B}_Y
  \;\simeq\;
  \frac{1}{\sigma_Y}\left[
    \mathbf{B}_Y\cdot(\nabla\times\mathbf{B}_Y)
    - \frac{ g'^2}{8\pi^2}\,
      \mu_{\text{eff}}^Y\,\mathbf{B}_Y^2
  \right].
  \label{EEBB}
\end{equation}
The time derivative of the hypermagnetic helicity density is given by~\cite{Grasso:2000wj}
\begin{equation}
  \frac{d h_Y}{dt}
  \;=\;
  -\,\frac{2}{V}\,
  \int_V d^3x\;
  \mathbf{E}_Y\cdot\mathbf{B}_Y.
  \label{eq:hdot-EB}
\end{equation}
Substituting Eq.~\eqref{EEBB} into Eq.~\eqref{eq:hdot-EB} yields the helicity evolution equation in terms of the effective chemical potential,
\begin{equation}
  \frac{d h_Y}{dt}
  \;\simeq\;
  -\,\frac{2}{\sigma_Y V}
  \int_V d^3x\left[
    \mathbf{B}_Y\cdot(\nabla\times\mathbf{B}_Y)
    - \frac{ g'^2}{8\pi^2}\,
      \mu_{\text{eff}}^Y\,\mathbf{B}_Y^2
  \right].
  \label{eq:hdot-final-mu}
\end{equation}
Eq.~\eqref{eq:hdot-final-mu} shows explicitly how the CPV effective chemical potential $\mu_{\text{eff}}^Y(t)$, associated with the time dependence of the Higgs modulus $\rho(t)$, biases the evolution of hypermagnetic helicity.

\begin{figure}[!htp]
\centering
\includegraphics[width=0.8\linewidth]{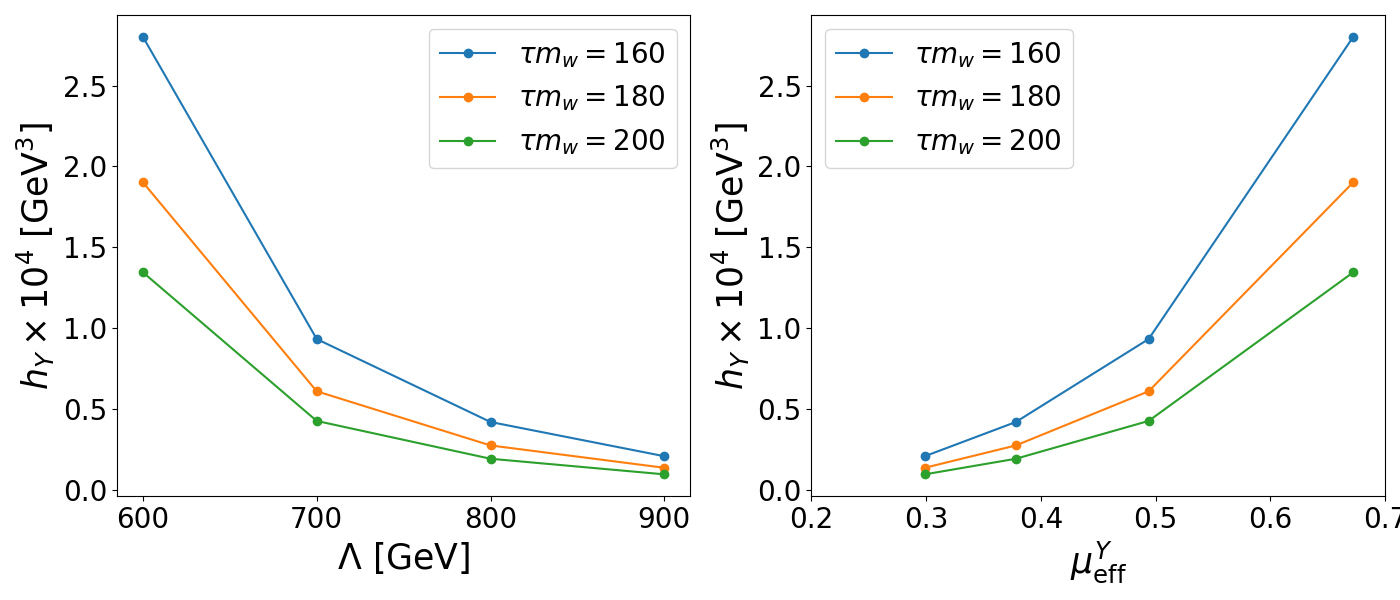}
\caption{Left: $h_Y$ as a function of the new physics scale $\Lambda$. Right: $h_Y$ as a function of the effective hypercharge chemical potential $\mu_{eff}^Y$. In both panels, the curves correspond to $\tau m_{\mathrm{w}}=160, 180, 200$, as indicated in the legend.}
\label{Hypermagnetic_density}
\end{figure}
After solving Eq.~\eqref{eq:hdot-final-mu} numerically, we obtain Fig.~\ref{Hypermagnetic_density}, which shows that the hypermagnetic helicity density $h_Y$ is sourced whenever the CPV effective chemical potential $\mu_{eff}^Y$, induced by the time-dependent Higgs modulus around the bubble walls, overlaps with a non-vanishing hypermagnetic field generated by the collision of bubbles. During the first-order EWPT, $\mu_{eff}^Y$ is localized in time around the period when the bubbles grow and collide, and the ring-like magnetic configuration produced by two expanding bubbles therefore acquires a non-vanishing Chern-Simons number~\cite{Ahonen:1997wh, Yang:2021uid}. Once the phase transition completes and $\mu_{eff}^Y\rightarrow0$, the source switches off and the generated hypermagnetic helicity is approximately frozen into the plasma, up to slow dissipative effects. The final value of $h_Y$ is thus controlled by the strength of the magnetic field created in the bubble collision and by the magnitude and duration of $\mu_{eff}^Y$, which in our setup are parametrized by the cutoff scale $\Lambda$ and the bubble wall time scale $\tau{m_{\mathrm{w}}}$.

The hypermagnetic helicity density is related to the $U(1)_Y$ Chern--Simons number density by~\cite{Giovannini:1997eg}:
\begin{equation}
  \Delta n_{\rm CS}^Y
  \;=\;
  \frac{g'^2}{16\pi^2}\,h_Y\;.
  \label{eq:nCS-hY-final}
\end{equation}
The resulting baryon-to-entropy ratio from this part at the end of the phase transition is
then~\cite{Kamada:2016eeb, Semikoz:2009ye}:
\begin{equation}
  \eta_B
  \;\equiv\;
  \frac{3\Delta n_{CS}^Y}{2s}
  \;\simeq\;
  \frac{g'^2}{16\pi^2}\,
  \frac{h_Y}{s}\,.
  \label{eq:etaB-hY-final}
\end{equation}

\begin{figure}[!htp]
\centering
\includegraphics[width=0.6\linewidth]{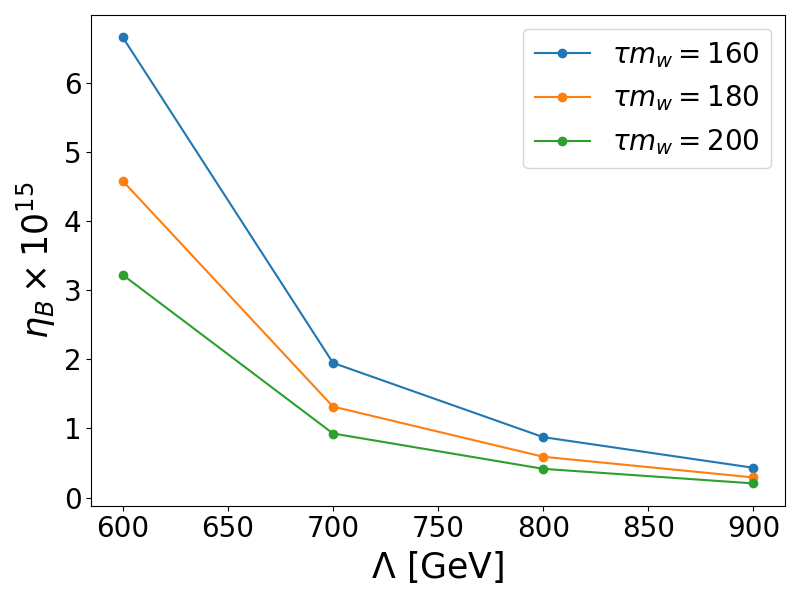}
\caption{
Dependence of the baryon-to-entropy ratio $\eta_B$ on the NP scale
$\Lambda$ for different values of the dimensionless bubble-wall collision times
$\tau m_w = 160,\,180,\,200$.  The points correspond to
$\Lambda = 600,\,700,\,800,\,900~\mathrm{GeV}$. }
\label{etablam}
\end{figure}

After solving Eq.~(\ref{eq:hdot-final-mu}) for the hypermagnetic helicity and converting it to $\eta_B$ via Eq.~(\ref{eq:etaB-hY-final}) at
$z{m_w}=0$, we present the resulting baryon asymmetry in Fig.~\ref{etablam}. The results show that it is hard to generate enough baryon number solely considering the chiral anomaly process during the first-order EWPT; therefore, the CME magnification effect illustrated in the previous section is crucial to interpret the observed BAU. 

We further note that the CPV operator $\tilde{O}_{\phi B}$ would further induce contributions to the electron electric dipole moment~\cite{Lue:1996pr}, 
\begin{equation}
\frac{d_e}{e}\approx\frac{m_e\sin^2\theta_W}{8 \pi^2\Lambda_{CPV}}\ln{\frac{\Lambda_{CPV}^2+m_H^2}{m_H^2}}\;.
\end{equation}
 Considering the SM Higgs mass $m_H=125$ GeV, the current constraints from the electric dipole moment\cite{Roussy:2022cmp}: $d_e/e<4.1\times 10^{-30} cm$ limits the CPV cutoff scale $\Lambda_{CPV}\gtrsim 338 $~TeV. The Fig.~\ref{etablam} demonstrates that such constraint rules out the possibility of the explanation of BAU with the effective chemical potential sourced from the current CPV operator. Next, following the Ref.~\cite{Garcia-Bellido:1999xos}, for the cutoff scale $\Lambda=338 \mathrm{TeV}$, the final baryon asymmetry can be obtained as $\eta_B=n_B/s=\frac{1}{s}\int dt \Gamma_{sph}\frac{\mu_{eff}}{T_{eff}}\simeq1.8\times10^{-14}$, which is still far below the observed value and therefore insufficient to account for the observed BAU. In the next section, we explore the situation when the CME effect is included.

\section{Baryogenesis with CME}

Recent numerical simulations suggest that the MF energy density is around the percent level of the radiation energy during the first-order EWPT~\cite{Zhang:2019vsb, Di:2020kbw}: $\rho_B\sim0.01\rho_{rad}$. 
Previous studies suggest the MF spectrum generated during the first-order EWPT should follow the form of $P_{(a)B}\sim 2\pi^2 B_\star^2 (k/k_\star)^{2(3)}$ when $k<k_\star$ for non-helical (helical) part~\cite{Durrer:2013pga} with the $B_\star$ being the MF magnitude at the correlation scale $k_\star=2\pi/R_\star$, here the $R_\star$ is the mean bubble separation~\cite{Huber:2008hg}. 
After generation, the joint evolution of the MF and the primordial plasma is governed by MHD. Once a magnetic field is generated, its subsequent evolution can be further affected by chiral effects. For definiteness, we focus on the electron sector and introduce independent chemical potentials, $\mu_L$ and $\mu_R$, for the approximately conserved number densities of left- and right-handed electrons, 
$n_{L,R} \equiv n_{e_{L,R}} = \mu_{L,R}T^2/6$. 
However, due to the chiral anomaly~\cite{Pavlovic:2016mxq} – a quantum effect leading to a change of the chiral electron number $n_L - n_R$ one has
\begin{equation}
    \begin{aligned}
        \frac{d(n_L - n_R)}{dt}
        = -\frac{2\alpha_Y}{\pi}\frac{1}{V}\int {\bf E_Y} \cdot {\bf B_Y}\, d^3x
        = \frac{\alpha_Y}{\pi}\frac{d\mathcal{H}}{dt}\;.
    \end{aligned}
\label{particle}
\end{equation}
Where, $\alpha_Y=g'^2/(4\pi)$ is the hypercharge fine-structure constant, $\mathcal{H}\equiv V^{-1}\int {\bf A_Y}\cdot{\bf B_Y}d^3x$ is the helicity density, and for chemical potentials $\mu_{L,R}=6n_{L,R}/T^2$. Therefore, we can rewrite Eq. (\ref{particle}) as
\begin{equation}
    \begin{aligned}
        &\frac{d \Delta\mu}{dt}=-\frac{c_{\Delta}\alpha}{T^2}\frac{d\mathcal{H}}{dt}\;,
    \end{aligned}
\label{del-mu}
\end{equation}
with $\Delta \mu=\mu_R-\mu_L$ being the difference between right and left chemical potentials, and the parameter $c_{\Delta}$ is a numerical coefficient of order unity that characterizes the dependence of the lepton number density $n_L$on the globally conserved charges in the primordial plasma.
The chiral anomaly leads to an additional contribution to the current in Maxwell's equations for the hypercharge fields, read
\begin{equation}
    \begin{aligned}
        &~\frac{\partial {\bf E_Y}}{\partial t}+{\bf J}+\frac{\alpha_Y\Delta\mu(t){\bf B_Y}}{\pi}=\nabla \times {\bf B_Y},~{\bf J} = \sigma_Y( {\bf E_Y}+{\bf v} \times {\bf B_Y}),\\
        &\frac{\partial{\bf B_Y}}{\partial t}=-\nabla \times {\bf E_Y},~\nabla \cdot {\bf E_Y}=0,~\nabla \cdot {\bf B_Y}=0\;.
    \end{aligned}
\label{amaxwell}
\end{equation}
 In this work, we neglect fluid advection {\bf v} and the displacement current, $\partial_t {\bf E_Y}$, which yields the CME-driven induction equation.
 \begin{equation}
     \begin{aligned}
         &\frac{\partial B_Y}{\partial t}=\frac{\alpha_Y}{\pi}\frac{\Delta \mu}{\sigma}\nabla \times B_Y+\frac{1}{\sigma}\nabla^2 B_Y.
     \end{aligned}
    \label{faraday}
 \end{equation}

 It is convenient to use conformal variables in a radiation-dominated Universe, with conformal time $\eta$, comoving wavenumber $k=a\tilde{k}$, and use the conductivity $\sigma_c=\sigma/T=100$.
  From Eq. (\ref{faraday}) we can obtain the evolution equations for the real binary products in the Fourier space, with $\rho_B(\eta)=\int dk \rho_k(k,\eta)=B_Y(\eta)^2/2$ is the magnetic energy density and $\mathcal{H}(\eta)=\int dk \mathcal{H}_k(k,\eta)$ is the magnetic helicity density, and satisfy the inequality $\rho_B(\eta) \geq k|\mathcal{H}_k(\eta)|/2$, which becomes saturated for field configurations termed maximally helical fields $\rho(k,\eta)=k |\mathcal{H}_k(k,\eta)|/2$ given by $\rho_B(\eta)=B_Y(\eta)^2/2$. 
 
 The general system of evolution equations for the spectra of the helicity density $\mathcal{H}_k(k,\eta)$ and the energy density $\rho_B(k,\eta)$ in the conformal coordinates reads~\cite{Boyarsky:2011uy}
 \begin{equation}
     \begin{aligned}
         &\frac{d \rho_B(k,\eta)}{d \eta}=-\frac{2k^2 }{\sigma_c}\rho_B(k,\eta)+\frac{\alpha_Y\Delta\mu k^2}{2 \pi\sigma_c}\mathcal{H}_k(k,\eta) ,\\
         &\frac{d \mathcal{H}_k(k,\eta)}{d \eta}=-\frac{2k^2 }{\sigma_c}\mathcal{H}_k(k,\eta)+\frac{2 \alpha_Y\Delta\mu}{\pi\sigma_c}\rho_B(k,\eta),\\
         &\frac{d \Delta \mu(\eta)}{d\eta}=-c_{\Delta}\alpha_Y\int dk \frac{\partial \mathcal{H}_k(k,\eta)}{\partial \eta}-\Gamma_f \Delta \mu(\eta)\;.
     \end{aligned}
 \label{speceq}
 \end{equation}
Where $\Gamma_f=10^{-2}h_e^2T/8\pi$ is the chirality flip rate and $h_e=2.8\times 10^{-6}$ is Yukawa coupling constant ~\cite{Abbaslu:2025ylq, Kamada:2016eeb},

The solution of the second equation in Eq. (\ref{speceq}) takes the form:
\begin{equation}
    \begin{aligned}
&\mathcal{H}_k(k,\eta)=\mathcal{H}_k(k,\eta_0)\exp(\frac{2k}{\sigma_c}(\frac{\alpha_Y}{2\pi}\int_{\eta_0}^{\eta}\Delta \mu(\eta^{\prime})d\eta^{\prime}-k(\eta-\eta_0))),
    \end{aligned}
    \label{helicalspec}
\end{equation}
$\mathcal{H}_k(k,\eta_0)$ is the helical magnetic field initial spectrum, and initial conditions at the time $\eta_0=7\times 10^{15}$ when the EWPT occurs. For convenience, one can use the notation of
\begin{equation}
    \begin{aligned}
        &\mathcal{H}_k(k,\eta)=\mathcal{H}_k(k,\eta_0)\exp(A_1(\eta)k-A_2(\eta)k^2)\;,\\
    \end{aligned}
    \label{mathcal}
\end{equation}
with
\begin{equation}
    \begin{aligned}    
        &A_1(\eta)=\frac{\alpha_Y}{\sigma_c\pi}\int_{\eta_0}^{\eta}\Delta \mu(\eta^{\prime})d\eta^{\prime},~A_2(\eta)=\frac{2}{\sigma_c}(\eta-\eta_0)\;.
    \end{aligned}
    \label{ABeta}
\end{equation}
The integrated helicity density is then
\begin{equation}
    \begin{aligned}
        &\mathcal{H}_k(\eta)=\int_{k_{min}}^{k_{max}}\mathcal{H}_k(k,\eta)dk\;.
    \end{aligned}
\label{heldensity}
\end{equation}

For a continuous initial spectrum $\mathcal{H}_k(k,\eta_0) = Ck^{n_s}=2\rho_B(k,\eta_0)/k$ with $n_s\geq 3$\cite{Semikoz:2013xkc,Semikoz:2015wsa,Durrer:2013pga}. 
The constant $C$ can be estimated using the relation for the full helical field. Normalizing to the initial MF amplitude $B_0\equiv B_Y(\eta_0)$ via $\int\rho_B(k,\eta_0)dk=B_0(\eta_0)^2/2$ gives
\begin{equation}
    \begin{aligned}
        &C=\frac{B_0^2(n_s+2)}{k_{max}^{n_s+2}-k_{min}^{n_s+2}}\;.
    \end{aligned}
    \label{Constant}
\end{equation}
For $k\rightarrow0$ violates the causal lower limit, $k>k_{min}=l_H^{-1}=H=1/2t\sim10^{-16}$ at $T\sim 100$ GeV, therefore, in the following calculations, we neglect $k_{min}$, where H is the Hubble rate. And, the $k_{max}$ is set by the correlation length, which corresponds to the mean bubble separation ($R_\star$) during the first-order EWPT, i.e., $k_{max}=2\pi/R_{\star}$~\cite{Huber:2008hg}. The $R_\star$ is tightly connected with the bubble number density as $R_{\star}\sim(1/n_b)^{1/3}$ with $n_b=8\pi \beta^3/v_w^3$~\cite{Hindmarsh:2019phv}. Here, the bubble wall velocity ($v_w$) and phase transition time ($\beta^{-1}$) are two crucial parameters to describe the EWPT. More explicitly, we have
\begin{equation}
    \begin{aligned}
k_{max}=\frac{2\pi(8\pi)^{1/3}\beta}{v_{\rm{w}}} \;.
    \label{kmax}
    \end{aligned}
\end{equation}

 Finally, the helicity density then evolves as~\cite{Semikoz:2013xkc}:
\begin{equation}
    \begin{aligned}
        &\mathcal{H}_k(\eta)=C\int_0^{k_{max}} k^{n_s}\exp(A_1(\eta)k-A_2(\eta)k^2)dk\;.
    \end{aligned}
\label{hel-den-con}
\end{equation}

We have the evolution of the asymmetry difference between left-handed and right-handed leptons $\Delta \xi_e(\eta)=\xi_{eR}(\eta)-\xi_{eL}(\eta)$, which is the dimensionless electron-asymmetry parameter for the chiral magnetic effect during the EWPT, $\Delta \xi_e(\eta)=\Delta\mu(\eta)/T$~\cite{Boyarsky:2011uy} The growth of $\Delta \xi_e(\eta)$ is due to the Abelian anomaly and its tendency to reach a constant value~\cite{Semikoz:2013xkc}. We here obtain a similar time independence of the saturation values of $\Delta \xi_e(\eta)$ for the continuous helicity density spectrum, we found the helicity density $\mathcal{H}_k(\eta)$ is almost conserved for the fully helical case at lower $k>k_{min}=l_H^{-1}=10^{-16}$. 

Then, we substitute the first and second equations in Eq.~(\ref{speceq}) into the third one with  maximally helical fields $\rho(k,\eta)=k |\mathcal{H}_k(k,\eta)|/2$, we have
\begin{equation}
    \begin{aligned}
        &\frac{d\Delta \mu}{d\eta}=c_{\Delta}\frac{2\alpha_Y}{\sigma_c} \int dk k^2 \mathcal{H}_k-c_{\Delta}\frac{\alpha_Y^2}{\pi\sigma_c}\int k\mathcal{H}_k\Delta \mu - \Gamma_f \Delta \mu\;.
    \end{aligned}
\end{equation}
This equation takes the standard form of a first-order linear differential equation.
\begin{equation}
    \begin{aligned}
        &\frac{d\Delta\mu}{d\eta}+R(\eta)\Delta \mu=S(\eta),
    \end{aligned}
    \label{SM}
\end{equation}
where
\begin{equation}
    \begin{aligned}
        &S(\eta)=c_{\Delta}\frac{2\alpha_Y}{\sigma_c} \int_0^{k_{max}} dk k^2 \mathcal{H}_k,~~~R(\eta)=c_{\Delta}\frac{\alpha_Y^2}{\pi\sigma_c}\int_0^{k_{max}} k\mathcal{H}_k + \Gamma_f \;.
        \label{SR001}
    \end{aligned}
\end{equation}
We substitute Eq.~(\ref{hel-den-con}) into Eq.~(\ref{SR001}) to obtain the following expression.
\begin{equation}
    \begin{aligned}
        &S(\eta)=\frac{2\alpha_Yc_{\Delta}}{\sigma_c}C\int_0^{k_{max}}k^{n_s+2}\exp[A_1(\eta)k-A_2(\eta)k^2]dk,\\
        &R(\eta)=\Gamma_f+\frac{\alpha_Y^2c_{\Delta}}{\pi\sigma_c}C\int_0^{k_{max}}k^{n_s+1}\exp[A_1(\eta)k-A_2(\eta)k^2]dk.
    \end{aligned}
\end{equation}
Integrating Eq.~(\ref{SM}), we obtain
\begin{equation}
    \begin{aligned}
        &\Delta \mu(\eta)=(\Delta\mu(\eta_0)-\frac{S(\eta)}{R(\eta)})\exp[-R(\eta)(\eta-\eta_0)]+\frac{S(\eta)}{R(\eta)}.
        \label{Deltamu001}
    \end{aligned}
\end{equation}
where $\Delta\mu(\eta_0)$ is the initial chiral anomaly. Since $\eta_0\sim10^{15}$, the first term on the right-hand side of Eq.~(\ref{Deltamu001}) can be approximated as zero. So $\Delta \mu(\eta)\approx S(\eta)/R(\eta)$. Defining
\begin{equation}
\begin{aligned} 
&I_m=\int_0^{k_{max}}k^{n_s+m}\exp[A_1k-A_2k^2]dk,m=1,2.
\end{aligned}
\end{equation}
we have
\begin{equation}
\begin{aligned}
    &I_2 \leq k_{max}I_1\;.
\end{aligned}
\end{equation}
Then
\begin{equation}
    \begin{aligned}
        &&\frac{S(\eta)}{R(\eta)}=\frac{\frac{2\alpha_Y}{\sigma_c}C I_2}{\Gamma_f+\frac{6\alpha_Y^2}{\pi \sigma_c}CI_1} \leq \frac{\frac{2\alpha_Y}{ \sigma_c}C I_2}{\frac{\alpha_Y^2}{\pi \sigma_c}CI_1}=\frac{2\pi}{\alpha_Y}k_{max} \Rightarrow \Delta \mu (\eta) \leq \frac{2\pi}{\alpha_Y}k_{max}.
    \end{aligned}
\end{equation}

After considering 't Hooft's conservation law $\eta_B(t)/3-L_e(t)=$ const ( $\eta_B=(n_B-n_{\Bar{B}})/s$ and $L_l=(n_l-n_{\Bar{l}})/s$ are the baryon and lepton numbers correspondingly), one obtain~\cite{Semikoz:2016lqv}:
\begin{equation}
    \begin{aligned}
        \frac{\partial(n_B-n_{\Bar{B}})/s}{\partial t}=\frac{3g'^2}{8\pi^2 s}{\bf E_Y}\cdot{\bf B_Y}\;.
    \end{aligned}
    \label{etaBB01}
\end{equation}
Combining with the third equation of Eq.~(\ref{speceq}), comparing the
result with Eq.~\ref{etaBB01}, and switching conformal variables gives
\begin{equation}
    \begin{aligned}
        &\eta_B(\eta)=\int_{\eta_0}^{\eta}-\frac{135g'^2}{32\pi^4 g_{\star} }\frac{d \mathcal{H}(\eta')}{d\eta'}d\eta'\;.
    \end{aligned}
    \label{etabb}
\end{equation}
From Eq.~(\ref{speceq}), Eq.~(\ref{hel-den-con}), Eq.~(\ref{etabb}), we have
\begin{align}
    \eta_B=\eta_B(\eta_0)+\frac{135g'^2}{32\pi^4g_{\star}}\mathcal{H}_k(k,\eta_0)(1-r),
    \label{etaBB}
\end{align}
with
\begin{equation}
\begin{aligned}
&r=\exp[\int_{\eta_0}^{\eta}\frac{-2k_{max}^2+\frac{\alpha_Y k_{max}}{\pi}\Delta(\eta')}{\sigma_c}d\eta']\;.
\end{aligned}
\end{equation}
Here, we consider the initial baryon number $\eta_B(\eta_0)\simeq0$.

Defining
\begin{equation}
\begin{aligned}  &\lambda(\eta')=\frac{-2k_{max}^2+\frac{\alpha_Y k_{max}}{\pi}\Delta(\eta')}{\sigma_c},\\
\end{aligned}
\end{equation}
one has
\begin{align}
    \lambda(\eta')=\frac{-2k_{max}^2+\frac{\alpha_Y k_{max}}{\pi}\Delta(\eta')}{\sigma_c}\leq \frac{-2k_{max}^2+\frac{\alpha_Y k_{max}}{\pi}\frac{2\pi}{\alpha}k_{max}}{\sigma_c}=0,
\end{align}
which implies $r\leq0$. Then according to Eq.~(\ref{speceq}), Eq.~(\ref{mathcal}), and Eq.~(\ref{etaBB}), $\eta_B\varpropto B_0^2$. 

\begin{figure}[!htp]
\centering
\includegraphics[width=0.4\linewidth]{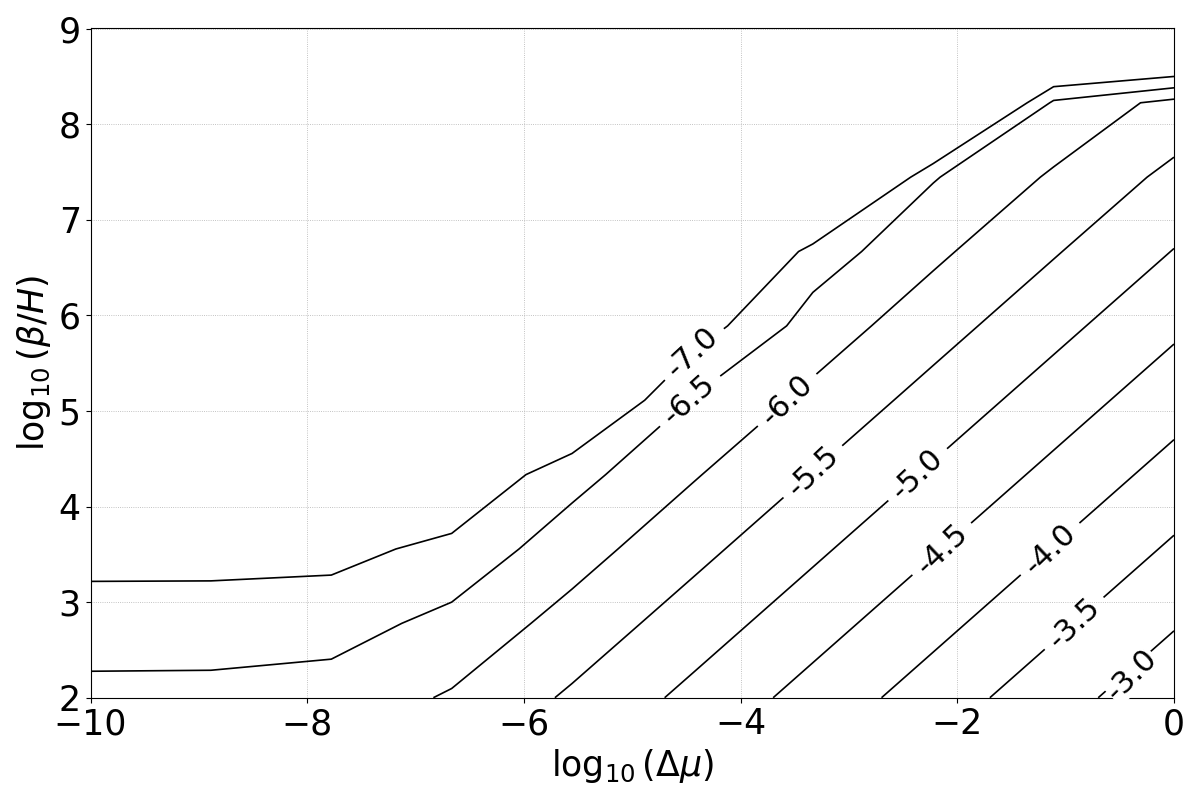}
\includegraphics[width=0.4\linewidth]{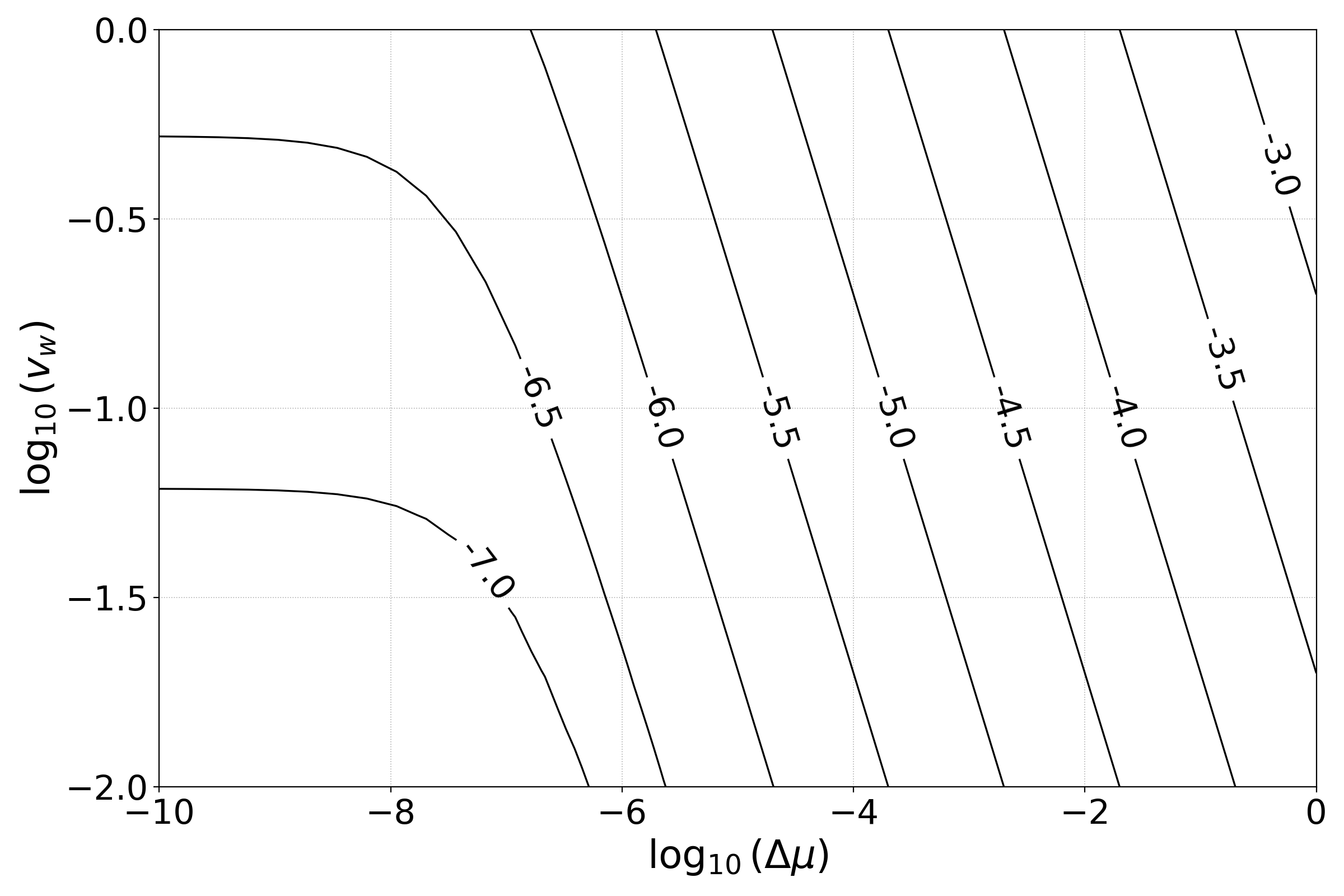}
\caption{Required initial magnetic field amplitude $B_0$ normalized by $m_{\rm w}^2$ to reproduce the observed BAU. Left:  $B_0/m_{\rm w}^2$ as a function of $\beta/H$ and $\Delta \mu$ with fixed $v_w=0.1$. Right: For $\beta/H=10^3$, the required $B_0$ versus $v_w$ and $\Delta \mu$.}  
\label{BAU_betaH}
\end{figure}

Fig.~\ref{BAU_betaH} summarizes how large the initial magnetic field $B_0$ must be (shown in units of $m_{\rm w}^2$) from the start to the completion of a first-order EWPT to reproduce the observed baryon asymmetry $\eta_B=8.8\times 10^{-11}$~\cite{Planck:2018vyg}. As depicted by the left panel, a faster transition enhances the conversion from helicity to baryons, so a smaller magnetic field is needed; hence $B_0$ decreases with $\beta/H$. A larger $\beta/H$ shifts the magnetogenesis spectrum to higher wavenumbers, increasing the helicity dissipation rate $\sim k^2/\sigma_c$; thus, a greater number of baryons is produced per unit of initial helicity $\mathcal{H}_k(\eta_0)$, and the required $B_0$ decreases. Increasing $\Delta \mu$ further enhances the chiral magnetic amplification $\propto k B \Delta \mu$, accelerating helicity processing, and again lowering the required $B_0$. As shown in the right panel, a larger wall velocity $v_w$ both strengthens the helical source and injects power at higher $k$, and a smaller $B_0$ suffices at larger $v_w$.

\section{Cosmological observation of the Magnetic field }

\begin{figure}[!htp]
\centering
\includegraphics[width=0.6\linewidth]{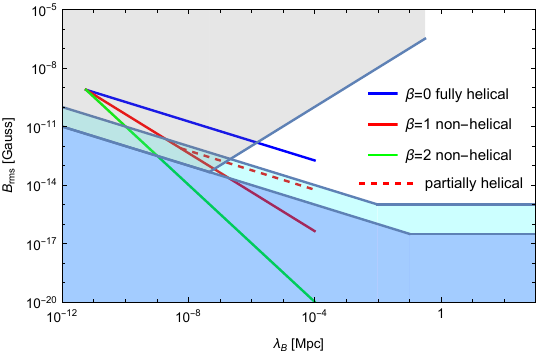}
\caption{$B_{rms}$ versus $\lambda_B$ for the fully helical case (blue), fractionally helical case (red dashed lines), and non-helical case (red and green). The gray region shows the upper bound on the MF  considering the MHD evolution~\cite{Taylor:2011bn, Banerjee:2004df}, and the cyan and blue regions are plotted to consider the bounds set by the Blazars~\cite{Fermi-LAT:2018jdy, Taylor:2011bn}. }\label{fig:bfcr}
\end{figure}

Here, we briefly discuss cosmic observations for these MFs generated during the EWPT. Since the phase transition type is of first-order, its correlation scale which is of the order of the size of mean bubble separation at coalescence, which is of the order of comoving correlation length $\lambda \simeq 0.01 l_H$~\cite{Durrer:2013pga}, $l_H$ denotes the comoving Hubble scale being given by~\cite{Kahniashvili:2012uj}: $l_H=5.8*10^{-10}{\rm Mpc}(100 {\rm GeV}/T)(100/g_\star)^{1/6}\;$,
Where $T$ is at the EWPT temperature. The physical magnetic field amplitude scales with the expansion of the universe as
$B_\star=(a_\star/a_0)^2B(T)$, where $B(T)$ is the MF strength from first-order EWPT as studied in the previous section, with the ratio of the scale factor at the time of MF generation to that at present being~\cite{Brandenburg:2017neh}: $a_\star/a_0 \simeq 8*10^{-16}(100~{\rm GeV}/T)(100/g_\star)^{1/3}$. The simulation of the evolution of hydromagnetic turbulence from the EWPT until today suggests that the root-mean-squared (nonhelical and helical) MF amplitude and the correlation length satisfy the following relation~\cite{Brandenburg:2017neh}:
$ B_{rms}=B_\star(\lambda_B/\lambda)^{-(\beta+1)/2}$.
Where $\beta=0$ is for the helical MF case, $\beta = 1,2$ for the non-helical MF case, and the fractionally helical MF case with $\epsilon_M=10^{-3}$~\cite{Brandenburg:2017neh}, and $\lambda_B$ is the correlation length.

The generated MF during the first-order EWPT under study will reach a fully helical case and cannot be probed currently; see Fig.~\ref{fig:bfcr} for illustration of the case $\Lambda=600$ GeV. Based on numerical simulations of the MFs generation from bubble collision during the first-order EWPT, Ref.~\cite{Di:2020kbw} also shows that the helical MF after evolution cannot be probed at present.

\section{Conclusion and discussion}

Using the SM model extended by dimension-six operators as a concrete example, we show that the two bubble collisions can generate the helical MF with a strength around $\sim \mathcal{O}(10^{-3}) ~m_W^2$. In contrast, the generated baryon asymmetry is far smaller than the observed BAU when one solely considers the Chern-Simons variation without including the CME effect. The electron electric dipole moment constraint makes it impossible to consider the CPV operator as the source of the effective chemical potential to trigger the baryon asymmetry generation.

We then perform a thorough study of the BAU in the background of helical MF. Our study builds a framework to address the BAU with CME during the first-order EWPT. We observe that to account for the observed the BAU, the phase transition parameters are required as: inverse duration $\beta/H\sim [100,10^8]$ and the bubble wall velocity $v_w\sim [10^{-2},1]$ for chemical potential of lepton $\Delta\mu\sim[\mathcal{O}(10^{-10}),\mathcal{O}(1)]$, wherein the MF strength should fall in the range of $B_0\sim[10^{-7},10^{-3}] m_W^2$. We then observe that the fully helical MF after its generation from the first-order EWPT cannot be probed in the near future.

In the study of first-order EWPT, the validity of the effective field theory with low cut-off scale $\Lambda$ has been questioned by Refs.~\cite{Damgaard:2015con, Postma:2020toi}, where the dimensional eight operators might be necessary~\cite{Chala:2018ari,Postma:2020toi,Hashino:2022ghd}. Refs.~\cite{Camargo-Molina:2021zgz, Camargo-Molina:2024sde} argued that the effective field theory can still be valid when the potential barrier necessitated by the first-order EWPT arises radiatively rather than from tree-level. 
Our conclusion would not be altered if effective field theory were invalidated, and the presence of a light degree of freedom (such as a new singlet or a second Higgs doublet) could accommodate relativistic bubble walls during the first-order EWPT. This is because our computations of BAU primarily rely on a strong MF generated by the collisions of expanding bubbles with a fast wall velocity. 

\section{Acknowledgements}

This work is supported by the National Key Research and Development Program of China under Grant No. 2021YFC2203004. L.B. is supported by the National Natural Science Foundation of China (NSFC) under Grants Nos. 12347101, 12322505. L.B. also acknowledges the Chongqing Natural Science Foundation under Grant No. CSTB2024NSCQ-JQX0022 and 
Chongqing Talents: Exceptional Young Talents Project No. cstc2024ycjh-bgzxm0020.

\appendix

\section{Equation of motions}
\label{sec:eom}
In this section, we derive EOMs relevant for the calculation of MFs production during the bubble collision process following Ref.~\cite{Stevens:2007ep}.
This Lagrangian under study is of the form:
\begin{align}
    &L_{EW} = L_1 + L_2 -V(\phi)+c_6\phi^{\dagger}\phi B_{\mu\nu}\tilde{B}^{\mu\nu}\;,
\label{lagrange}
\end{align}
with
\begin{align}
    &L_1 = -\frac{1}{4}W_{\mu\nu}^{i}W^{i\mu\nu}-\frac{1}{4}B_{\mu\nu}^{i}B^{i\mu\nu}\;, \nonumber\\
    &W_{\mu\nu}^{i} = \partial _\mu W_\nu^{i}-\partial _\nu W_\nu^{i}-g \epsilon_{ijk}W_{\mu}^{j}W_{\nu}^{k} \;,\nonumber \\
    &B_{\mu\nu}=\partial _\mu B_\nu-\partial _\nu B_\nu \;,\nonumber \\
    &L_2 = |i\partial _\mu-\frac{g}{2}\tau\cdot W_{\mu}-\frac{g^{\prime}}{2}B_{\mu}\Phi|\;,\\
    &V(\phi)=\mu_h^2\Phi^\dag \Phi+\lambda (\Phi^\dag \Phi)^2+c_6 (\Phi^\dag \Phi)^3\;.\nonumber
\end{align}
Where the $W^{i}$, with ${i=(1,2)}$, are the $W^{\pm}$fields, the $U(1)_Y$ hypercharge gauge field $B_{\mu}$, $\Phi$ is the Higgs field, $\tau^{i}$is the SU(2) generator, $B_{\mu\nu}$ is electromagnetic field strength tensor, $c_6=\Lambda^{-2}$ and $\tilde{B}^{\mu\nu}$ is the dual tensor, and $\Lambda$ is the energy scale of new physics that suppresses the dimension-six CP-violating operator. 
The electromagnetic field strength tensor is given:
\[
B^{\mu\nu}=\begin{pmatrix}
    0&E_x&E_y&E_z&\\
    -E_x&0&-B_z&-B_y&\\
    -E_y&-B_z&0&B_x&\\
    -E_z&B_y&-B_z&0&
\end{pmatrix}\;,
\]
The electromagnetic and Z fields are defined as
\begin{align}
    &A_{\mu}^{EM}= \frac{1}{\sqrt{g^{\prime 2}+g^{2}}}(g^{\prime} W_{\mu}^{3}+gB_{\mu}) \;, \\
    &Z_{\mu}= \frac{1}{\sqrt{g^{\prime 2}+g^{2}}}(g W_{\mu}^{3}-g^{\prime}B_{\mu})\;.
\end{align}
Here, $g=\frac{e}{\sin(\theta_w)}=0.646, g^{\prime}=g\tan(\theta_w)=0.343$, and the paper units are such that $\hbar=c=1$. Equations of motion from the Lagrangian given by Eq. (\ref{lagrange}) can be obtained by applying the principle of least action $\delta\int d^4xL_{EW}=0$. The modulus $\rho$ of the Higgs field satisfies the $\rho$-equation
\begin{equation}
    \begin{aligned}
        0&=\partial^2\rho(x)-\frac{g^2}{4}\rho(x)[{ W}^1\cdot{ W}^1+{ W}^2\cdot{ W}^2]
        -\rho(x)\psi_{\nu}\psi^{\nu}+\rho(x)\frac{\partial V}{\partial \rho^2}-\frac{\rho(x) B_{\mu\nu} \tilde{B}^{\nu\mu}}{\Lambda^2}\;.
    \end{aligned}
\end{equation}
Where the quantity $\psi_{\nu}$ is defined in terms of the phase of the Higgs field and the Z field.
The B field satisfies:
\beq
\label{L2}
    \partial^2B_\nu-\partial_\nu\partial\cdot B+g'\rho(x)^2\psi_\nu(x) = \frac{4 \partial^{\mu}(\rho(x)^2) \tilde{B}_{\nu\mu}}{\Lambda^2}\;,
\eeq
where the $\psi_\nu$ is
\begin{equation}
\begin{aligned}
&\psi_\nu(x)\equiv \partial_\nu \Theta - \frac{\sqrt{g^2+g'^2}}{2}Z_\nu\;, 
\end{aligned}
\end{equation}
and satisfies
\beq\label{eqpsi}
\partial^{\nu}\left(\rho(x)^2\psi_\nu(x)\right)=0\;.
\eeq
For $i=3$, gauge field $W^i$ satisfies the following $W$-equation
\beq
\partial^2W^3_\nu-\partial_\nu\partial\cdot W^3-g\rho(x)^2\psi_\nu(x)=j^{3}_\nu(x)\;,
\eeq
and, for $i=1,2$, we have
\beq
\partial^2W^i_\nu-\partial_\nu\partial\cdot W^i+m_W(x)^2W^i_\nu=j^{i}_\nu(x)\;,
\eeq
with $m_W(x)^2=g^2\rho(x)^2/2$,
and $j^i_\nu(x)$ is,
\beq
j^i_\nu(x) \equiv g\epsilon_{ijk}(W^k_\nu\partial\cdot W^j +2W^j\cdot\partial W^k_\nu-W^j_\mu\partial_\nu W^{k\mu})
           - g^2\epsilon_{klm}\epsilon_{ijk}W^j_\mu W^{l\mu}W^m_\nu\;.
\eeq
The EOM for $A^{em}$ casts the form of,
\beq\label{eqA}
\partial^2A^{em}_\nu-\partial_\nu\partial \cdot A^{em}=j^{em}_\nu(x)\;,
\eeq
with
\begin{equation}
\begin{aligned}
&c_1=\frac{g'}{\sqrt{g^2+g'^2}},~~c_2=\frac{g }{\sqrt{g^2+g'^2}} \\
&j^{em}_\nu(x)=c_1 \times j^3_\nu(x)+c_2 \times \frac{4 \partial^{\mu}(\rho(x)^2) \tilde{B}_{\nu\mu}}{\Lambda^2}\; .
\end{aligned}
\end{equation}
And, the EOM for the $Z$ field is obtained as,
\beq\label{eqZ}
\partial^2Z_\nu-\partial_\nu\partial\cdot Z-\rho(x)^2\sqrt{g^2+g'^2}\psi_\nu(x)
-c_2 \times j_{\nu}^3+\frac{4c_1}{\Lambda^2} \partial^{\mu}(\rho(x)^2)\tilde{B}_{\nu\mu}=0\;.
\eeq

Utilizing the thermal erasure~\cite{Stevens:2012zz} of $\langle Z\rangle=0$. Applying the ensemble averaging to Eq.~(\ref{eqA}), we can obtain
\beq
\langle j^{em}_\nu\rangle=(\frac{c_1}{g}+c_2)\frac{4 \partial^{\mu}(\rho(x)^2) \tilde{B}_{\nu\mu}}{\Lambda^2}-\frac{g'}{c_2}\rho{(x)}^2\times \partial_\nu\Theta(x)\;
\eeq

Consequently, the Eq. (\ref{eqA}) recasts the form of the Maxwell equation,
\begin{equation}
\begin{aligned}
&\partial^2A_\nu-\partial_\nu\partial\cdot A =j^{em}_\nu(x)=j_1(x)+j_2(x)
\label{jem}
\end{aligned}
\end{equation}
where 
     % \[
     %    j_1(x) =(\frac{c_1}{g}+c_2)\frac{4 \partial^{\mu}(\rho(x)^2)\tilde{B}_{\nu\mu}}{\Lambda^2}=
     %    \begin{cases} 
     %    -(\frac{c_1}{g}+c_2)\frac{4(\nabla \rho(x)^2)\cdot {\bf B}}{\Lambda^2}, & \nu=0, \\ 
     %    (\frac{c_1}{g}+c_2)\frac{8\rho(x)(\frac{\partial \rho(x)}{\partial t}{\bf B}-(\nabla\rho(x)) \times {\bf{E} })}{\Lambda^2}}, & \nu = 1,2,3\;,\\
     %    \end{cases}
     %    \]}
     \[
    j_1(x) = \left(\frac{c_1}{g} + c_2\right) \frac{4 \partial^{\mu}(\rho(x)^2) \tilde{B}_{\nu\mu}}{\Lambda^2} =
    \begin{cases}
    -\left(\frac{c_1}{g} + c_2\right) \frac{4 (\nabla \rho(x)^2) \cdot {\bf B}}{\Lambda^2}, & \nu = 0, \\
    \left(\frac{c_1}{g} + c_2\right) \frac{8 \rho(x) \left(\frac{\partial \rho(x)}{\partial t} {\bf B} - (\nabla \rho(x)) \times {\bf E}\right)}{\Lambda^2}, & \nu = 1, 2, 3.
    \end{cases}
\]
        and
\begin{equation}
    \begin{aligned}
        &j_2(x)=-\frac{g'}{c_2}\rho{(x)}^2\times \partial_\nu\Theta(x)\nonumber\;.
    \end{aligned}
\end{equation}

Specifically, the magnitudes of ${\bf B}$ are determined by the contribution of the Higgs phase $\theta(x)$ from the current $j_2(x)$. When the Higgs fields of the colliding bubbles differ in phase before the collision, $\partial_\nu\Theta(x) $ develops a non-zero value within the bubble overlap region after the collision. The magnetic field strength can be calculated using ${\bf B}=\nabla \times {\bf A}$ after obtaining  the electromagnetic current, i.e.,
 \begin{equation}\label{eqB}
     \begin{aligned}
         &\nabla^2 {\bf B}=\nabla^2(\nabla \times {\bf A^{em}})=\nabla \times (\nabla^2{\bf A^{em}})=-\nabla \times { \bf  j^{em}}\;.
     \end{aligned}
 \end{equation}
The electromagnetic current $j^{em}_{\nu}(x)$ then develops a non-zero value there and creates a magnetic field in units of $m_{\rm w}^2$ through Eq. (\ref{jem}, \ref{eqB}).

\section{Phase transition and the induced Chern-Simons term}
\label{sec:csd}

We study the Chern-Simons density in the Standard Model Effective Theory with the dimensional reduction (DR) in this part. The original $4d$ Lagrangian is given in Eq.~(\ref{lagrange}).
 The operator $\Phi^{\dag}\Phi B_{\mu\nu}\tilde{B}^{\mu\nu}$ will change the wave function renormalization, but this change is at order $\mathcal{O}(g^4)$ due to the power counting $c_6\sim g^4/\Lambda^2$~\cite{Qin:2024idc}. This leads to a higher order than $\mathcal{O}(g^4)$ in the 3d coupling matching, and it should be ignored in the study of phase transition dynamics. After integrating out the super heavy mode, the effective theory has the form~\cite{Qin:2024idc, Gynther:2003za}
\begin{equation}\label{action3dheavy}
\mathcal{L}^{heavy}_{3d} =
      \frac{1}{4}W_{ij}^{a}W_{ij}^{a}
    + \frac{1}{4}B_{ij}^{ }B_{ij}^{ }
    +\frac{1}{2}(D_{i}^{ }W_{0}^{a})^2
    + \frac{1}{2}(\partial_{i}^{ }B_{0}^{ })^2
    + \frac{1}{2}(D_{i}^{ }C_{0}^{\alpha})^2
    + (D_{i}\Phi^{ })^\dagger (D_{i}\Phi^{ })
    + V^{heavy}_{3d}
    \;,
\end{equation}

where
$W^{a}_{ij} =
\partial_{i} W^a_j -
\partial_{j} W^a_i + g_{3} \epsilon^{abc}W_{i}^b W_{j}^c$,
$B_{ij} = \partial_{i}B_{j} - \partial_{j}B_{i}$ and
$D_{i}\Phi = (\partial_{i} - ig_{3} \tau^a W_{i}^a/2 - ig^\prime_{3} B^{ }_{i}/2)\Phi^{ }$ with $\tau^a$ being the Pauli matrices. The heavy scalar potential has the form
\begin{equation}\label{heavypotential}
\begin{aligned}
V^{heavy}_{3d} =&
    \mu_{h,3}^{2} \Phi^\dagger\Phi
    + \lambda^{ }_{3} (\Phi^\dagger\Phi)^2
    + c^{ }_{6,3} (\Phi^\dagger\Phi)^3
    + \frac{1}{2}m_D^2\,W^a_{0}W^a_{0}
    + \frac{1}{2}m_D'^2\,B_{0}^2
    + \frac{1}{2}m_D''^2\,C^{\alpha}_{0} C^{\alpha}_{0}
    \\ &
    + \frac{1}{4}\kappa_1 (W^a_{0}W^a_{0})^2
    + \frac{1}{4}\kappa_2 B_{0}^4
    + \frac{1}{4}\kappa_3 W^a_{0}W^a_{0}B_{0}^2
    + h^{ }_{1} \Phi^\dagger\Phi W^a_{0}W^a_{0}
    + h^{ }_{2} \Phi^\dagger\Phi B_0^2+\kappa_0 B_{0}
    \\&
    + h^{ }_{3} B_{0}\Phi^\dagger{W}_{0}^a\tau^a\Phi
    + h^{ }_{4} \Phi^\dagger\Phi C^\alpha_{0} C^\alpha_{0}
    +\rho \Phi^\dagger W_{0}^a\tau^a \Phi+\rho^\prime \Phi^\dagger \Phi B_{0}+\rho_G B_{0} W_{0}^a W_{0}^a
    \\&
     +\alpha_0 \epsilon_{ijk}\left(W_{i}^a W_{jk}^a-\frac{i}{3}g_3\epsilon^{abc} W_{i}^a W_{i}^b W_{k}^c\right)+\alpha^\prime\epsilon_{ijk} B_{i} B_{jk}
    \;,
\end{aligned}
\end{equation}
Where $m_D,m_D',m_D''$ are the Debye masses of the time component of gauge fields. The parameters of heavy scalar potential Eq.~\eqref{heavypotential} are defined in Ref.~\cite{Qin:2024idc}. The Chern-Simons terms $\alpha_0$ and $\alpha^\prime$ are defined as
\begin{align}\label{NCS}
\alpha_0=\frac{g^2}{32\pi^2}(N_f \mu_B+\sum_{i=1}^3\mu_{L_i}),\quad \alpha^\prime=&-\frac{g^{\prime 2}}{32\pi^2}(N_f \mu_B+\sum_{i=1}^3\mu_{L_i})\;,
\end{align}
where the $N_f$ is the number of families, and $\mu_B,\mu_{L_i}$ are the baryon and lepton chemical potential respectively. The relation $(N_f \mu_B+\sum_{i=1}^3\mu_{L_i})$ can be obtained by assuming the left- and right-hand fermions have the same chemical potential, and it vanishes by defining~\cite{Qin:2024idc}~\cite{Gynther:2003za}
\begin{equation}\label{BLasum}
\mu_B=\frac{1}{N_f}\sum_{i=1}^{N_f}\mu_i,\quad \mu_{L_i}=-\mu_i\;.
\end{equation}

However, the lepton chemical potential is not equal to each other($\mu_{L_i}\neq\mu_{R_i}$) since the lepton asymmetry, where $\mu_{L_i}$ is the chemical potential of left-hand lepton and $\mu_{R_i}$ is the chemical potential of right-hand lepton. Then the Chern-Simons term with the lepton asymmetry has the form~\cite{Laine:2005bt}
\begin{align}\label{NCSnew}
c_E \epsilon_{ijk}\biggr(W_{i}^a W_{jk}^a-&\frac{i}{3}g_3\epsilon^{abc} W_{i}^a W_{i}^b W_{k}^c\biggr)+c_E^\prime\epsilon_{ijk} B_{i} B_{jk},\\
c_E= 3N_f\mu_B+\sum_{i=1}^{N_f}&\mu_{L_i},\quad c_E^\prime=-c_E+2\sum_{i=1}^{N_f}(\mu_{L_i}-\mu_{R_i}).
\end{align}
The rate of $B+L$ violation is significantly larger than the Hubble rate, and the system can pass over the barrier between the different vacua instead of penetrating through the barrier since the rate of the anomalous non-conservation of the fermion number can be unsuppressed in this system~\cite{Laine:2005bt, Kuzmin:1985mm}. For this reason, the abnormal processes are perfectly in thermal equilibrium, and the chemical potential should be set to zero $\mu_{B+L}=3N_f\mu_B+\sum_{i=1}^{N_f}\mu_{L_i}=0$ and $c_E$ vanishes.

But the $c_E^{\prime}$ does not vanish because the right-hand lepton($e_R$) is coupled with the $U(1)_Y$ hypercharge gauge field. This interaction does not change the quantum numbers or generation, while the Yukawa interaction can change the quantum numbers or generation. The Yukawa interaction between the Higgs and electron is very weak for the tiny electron mass. That leads to the $e_R$ coming into chemical equilibrium at temperature $T\simeq 1$ TeV. Since the sphaleron process falls out of equilibrium near this temperature,  the $e_R$ may not be transformed into $e_L$ soon enough, and the initial $e_R$ asymmetry is not washed out~\cite{Campbell:1992jd, Cline:1993vv}.

\begin{figure}[!htp]
\centering
\includegraphics[width=0.4\linewidth]{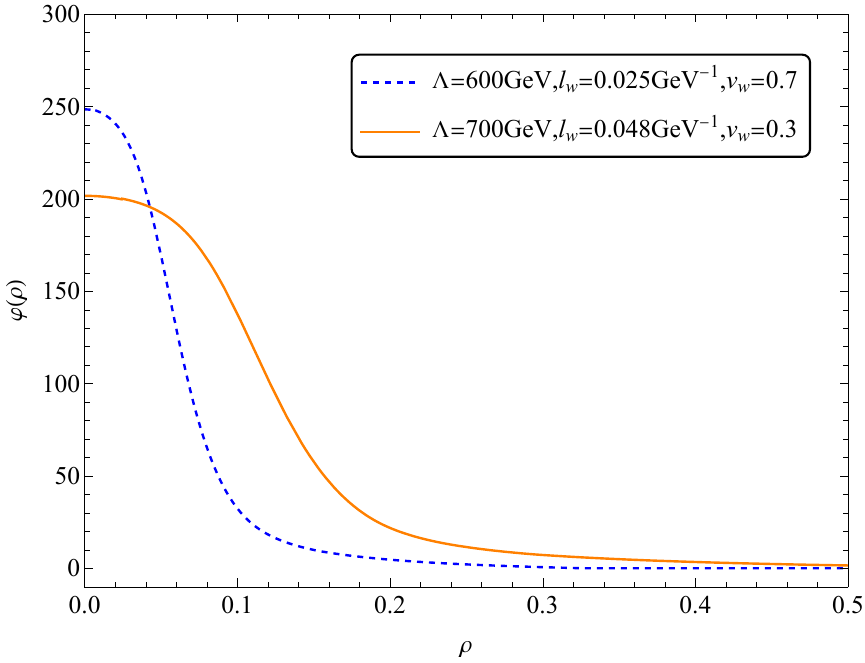}
\label{Fig. sub.2}
\includegraphics[width=0.4\linewidth]{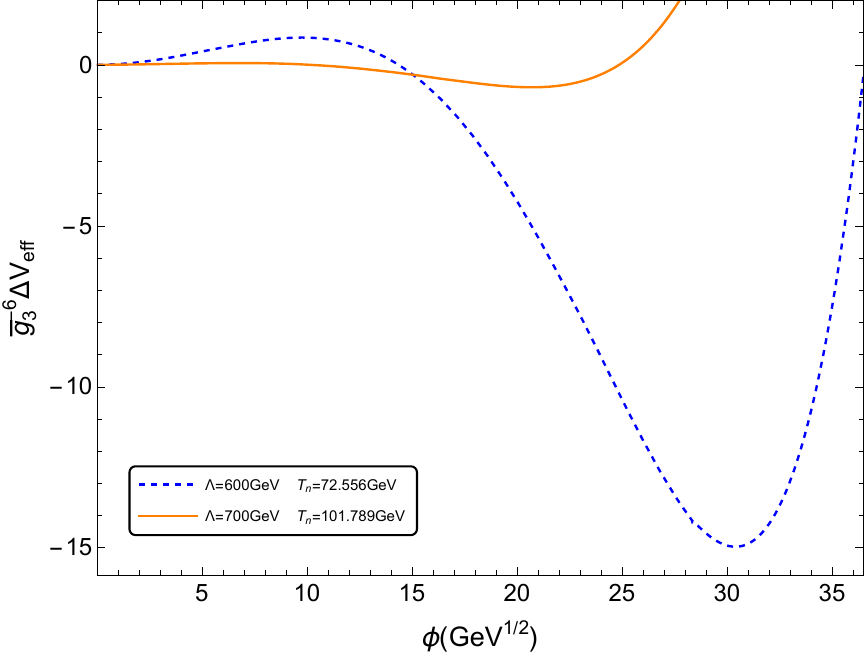}
\caption{The initial bubble profiles (left) and the corresponding effective thermal potential at the nucleation temperature 
(right). }
\label{phi}
\end{figure}

The Fig.~\ref{phi} shows the effective potential $\Delta V_{eff}$ and bounce profile at temperature $T_n$ with the $\Lambda=600$ GeV and $700$ GeV. The  $\Delta V_{eff}=V_{eff}(\phi)-V_{eff}(0)$ is calculated in light scale and has dimension GeV$^3$, and $\overline{g}_3^2$ is the $SU(2)$ gauge coupling in the light scale with dimension GeV$^2$~\cite{Qin:2024idc}. The nucleation temperature $T_n$ is obtained when the bubble nucleation rate $\Gamma=A \exp[-S_c]$ is equal to Hubble parameter $\Gamma\sim H$, i.e., $S_c\approx 140$~\cite{Linde:1981zj}.
 The Euclidean action is
\begin{equation}\label{action}
S_c=\int \bigg[\frac{1}{2}(\partial_i \phi)^2+V_{eff}(\phi,T) \bigg]d^3 x.
\end{equation}
The action can be obtained by solving the bounce function:
\begin{equation}\label{bouncefun}
\frac{d^2\phi}{d\rho^2}+\frac{2}{\rho}\frac{d\phi}{d\rho}=\frac{dV_{eff}(\phi,T)}{d\phi}\;,
\end{equation}
With the boundary condition
\begin{equation}
\phi(\rho\rightarrow \infty)=0,\quad \left.\frac{d\phi}{d\rho}\right|_{\rho=0}=0\;,
\end{equation}
and we use code ``findbounce'' to solve this equation and obtain the nucleation temperature $T_n$ and the corresponding background field value $\phi_n$~\cite{Guada:2020xnz}. The inverse PT duration is defined as: $\beta/H_n=T_n(dS_c/dT)|_{T_n}$. The PT temperature and the duration determine the peak frequency of the produced GW from PT\cite{Huber:2008hg,Caprini:2015zlo}, and the trace anomaly ($\alpha$) usually determines the amplitude of the generated GW. For the 4d theory, the $\alpha$ is defined as $\alpha=\Delta\rho/\rho_{rad}$ with 
\begin{equation}
    \begin{aligned}
        \Delta\rho=\Delta V_{4d}(\phi_n,T_n)+\frac{1}{4}T_n\frac{d\Delta V_{4d}(\phi_n,T)}{dT}|_{T=T_n}
    \end{aligned}
\end{equation}
After apply the relation between 4d and 3d potential $V_{4d}\approx TV_{eff}$, we have $\alpha=T(\Delta\rho/\rho_{rad})$ with
\begin{equation}
    \begin{aligned}
         \Delta\rho=-\frac{3}{4}\Delta V_{eff}(\phi_n,T_n)+\frac{1}{4}T_n\frac{d\Delta V_{4d}(\phi_n,T)}{dT}|_{T=T_n}
    \end{aligned}
\end{equation} and $\rho_{rad}=\pi^2 g_{\star}T^4/30$, $g_{\star}$ is the effective number of relativistic degrees of freedom~\cite{Caprini:2009yp}.
where $\Delta V_{eff}(\phi,T)=V_{eff}(\phi,T)-V_{eff}(0,T)$. 
The $l_w$ in SMEFT has the form~\cite{Lewicki:2021pgr}
\begin{equation}\label{lw}
  l_w=\sqrt{\frac{\phi_n^2}{4V_h}},\quad V_h=\Delta V_{eff}^{max}-\Delta V_{eff}^{min},
\end{equation}
where $V_h$ is the height of the potential barrier. The bubble wall velocity is defined as~\cite{Dine:1990fj}
\begin{equation}\label{vw}
v_w=\begin{cases}
    \sqrt{\frac{\Delta V}{\alpha_n \rho_r}}&, \quad \sqrt{\frac{\Delta V}{\alpha_n \rho_r}}<v_J(\alpha_n)\\
    1&,\quad \sqrt{\frac{\Delta V}{\alpha_n \rho_r}}\geq v_J(\alpha_n)
\end{cases}
\end{equation}
with the $\Delta V$ being the difference between the broken phase and symmetric phase, and the Jouguet velocity $v_J(\alpha_n)$ is~\cite{Lewicki:2021pgr, Dine:1990fj}
\begin{equation}
v_J=\frac{1}{\sqrt{3}}\frac{1+\sqrt{3\alpha_n^2+2\alpha_n}}{1+\alpha_n}\;,
\end{equation}
where $\alpha_n$ is the strength of the phase transition.

\bibliography{references}

\end{document}